\documentclass[aps,preprint,groupedaddress,showpacs]{revtex4}

\usepackage{graphicx}
\usepackage{amsfonts,amssymb,amsmath,bm}

\begin{document}


\title{Multiband superconductivity in the heavy fermion compound PrOs$_{4}$Sb$_{12}$}



\author{ G. Seyfarth$^{1}$, J.P. Brison$^{1}$, M.-A. M{\'e}asson$^{2}$, J. Flouquet$^{2}$, K. Izawa$^{2,3}$, Y. Matsuda$^{3,4}$, H. Sugawara$^{5,6}$, and H. Sato$^{5}$}

\affiliation{$^{1}$CRTBT, CNRS, 25 avenue des Martyrs, BP166, 38042 Grenoble CEDEX 9, France}
\affiliation{$^{2}$DRFMC,SPSMS, CEA Grenoble, 38054 Grenoble, France}
\affiliation{$^{3}$Institute for Solid State Physics, University of Tokyo, Kashiwanoha, Kashiwa, Chiba 277-8581, Japan}
\affiliation{$^{4}$Department of Physics, Kyoto University, Sakyo-ku, Kyoto 606-8502, Japan}
\affiliation{$^{5}$Department of Physics, Tokyo Metropolitan University, Hashioji, Tokyo 192-0397, Japan}
\affiliation{$^{6}$The University of Tokushima, Minamijosanjima-machi, Tokushima 770-8502, Japan}


\date{\today}

\begin{abstract}
The thermal conductivity of the heavy fermion superconductor
PrOs$_{4}$Sb$_{12}$  was measured down to T$_{c}$/40 throughout the
vortex state. At lowest temperatures and for magnetic fields
$H\approx0.07H_{c2}$, already $40\%$ of the normal state thermal
conductivity is restored. This behaviour (similar to that
observed in MgB$_{2}$) is a clear signature of multiband
superconductivity in this compound.

\end{abstract}

\maketitle


The low temperature properties of PrOs$_{4}$Sb$_{12}$, the first Pr-based heavy fermion (HF) superconductor \cite{Bauer I} (filled skutterudite structure with space group Im$\overline{3}$,
$T_{c}\simeq$1.85~K) have many unusual features both in the normal and in the superconducting states \cite{Aoki2005}: the non-magnetic singlet ground state of the Pr$^{3+}$ ion suggests that the conduction electron mass renormalization comes from inelastic scattering by crystal field transitions, whereas the superconducting transition temperature would be enhanced by the quadrupolar degrees of freedom of the rare earth $f$ electrons \cite{Goremychkin, Kuwahara2004}. Several experiments also point to \textit{unconventional} superconductivity in this compound: a double superconducting transition in the specific heat \cite{Maple, Vollmer, Marie-Aude} as well as thermal conductivity measurements in a rotated magnetic field \cite{Koichi} could result from different symmetry states of the order parameter; London penetration depth studies \cite{Chia} or flux-line lattice distortion \cite{Huxley2004} indicate nodes of the gap. Tunneling spectroscopy reveals a gap of the order of the BCS value, but with a distribution of values as observed in borocarbides or in NbSe$_2$ \cite{Hermann2004}.

In this Letter, we report a study of heat transport at very low temperature and under magnetic field in PrOs$_{4}$Sb$_{12}$, intended to probe the low energy excitations, i.e. the gap structure and nodes. Instead, another phenomenon was uncovered: our results provide compelling evidence for multiband superconductivity (MBSC) in this compound.
Coming after similar findings in MgB$_{2}$\cite{Bouquet2001}, NbSe$_2$  \cite{Boaknin2003} or in the borocarbides Y and LuNi$_2$B$_2$C \cite{Shulga}, they show that very diverse mechanisms may lead to MBSC (or strongly anisotropic gaps) so that it could be much more common than presently thought.

Our rectangular-shaped ($\sim0.4\times0.4\times2$~mm$^{3}$) PrOs$_{4}$Sb$_{12}$  single crystal (same as in \cite{Koichi}) was grown by the Sb-flux method \cite{Koichi} and has $T_{c}\simeq1.85$~K . The thermal conductivity ($\kappa$) parallel to the magnetic field
was measured in a dilution refrigerator by a standard two-thermometers-one heater steady-state method down to 50~mK and up to 2.5~T ($\mu_{0}H_{c2}(T\longrightarrow0)\simeq2.2$~T).  The carbon thermometers were 
thermalized on the sample by gold wires, spot welded on the surface of the PrOs$_{4}$Sb$_{12}$ sample. 
The same contacts and gold wires were used to measure the electric resistivity of the sample by a standard four-point lock-in technique.

\begin{figure}[t]
 \includegraphics[width= 15cm]{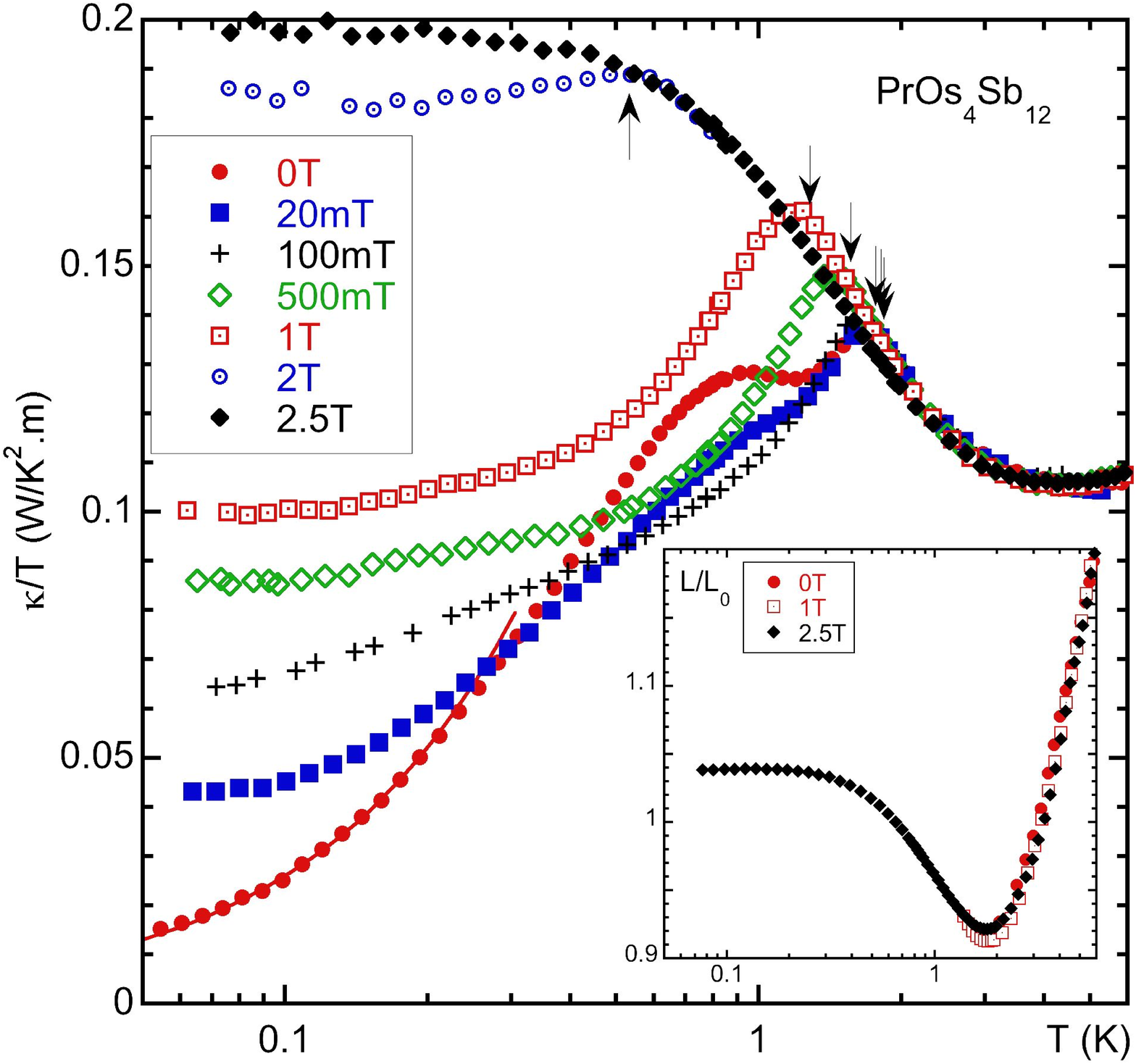}
 \caption{$\kappa(T)/T$ at different fields. The anomaly at $T\approx$ 1~K is rapidly suppressed under magnetic fields. Full line: pure $\kappa/T=aT$ law valid in zero field below 0.2~K.  Arrows: $T_c(H)$. Insert: test of the Wiedemann-Franz law. Color online.
 }
 \label{kappa_all}
 \end{figure}

Figure \ref{kappa_all} shows the temperature dependence of
$\kappa/T$ at different magnetic fields. Defining $L=\kappa\rho/T$, the insert demonstrate the excellent agreement (within 3~$\%$) with the Wiedemann-Franz law at the lowest temperatures in the normal state (data in 2.5~T). The minimum of $L/L_0$ at temperatures around 1~K reveals the growth of inelastic collisions on warming. For $T\geq3$~K, $L/L_0$ increases above 1, maybe due to a phonon contribution to the heat transport (about 20~$\%$ at 6~K).

At the superconducting transition, $\kappa/T$ (zero field data) exhibits no anomaly, as predicted by ordinary BCS theory. The decrease of $\kappa(T,H=0)$ seems to take place only slightly below $T_{c}$, when the number of excited quasiparticles is reduced by the gap opening. In our case, $\kappa/T$ exhibits a significant enhancement at around $T_c/2\approx$ 1~K. This feature is suppressed by a field of only 20~mT (see figure \ref{kappa_all}), whereas the specific heat remains unchanged under such small magnetic fields (results not shown here). So this anomaly should be controlled by the scattering mechanism. One possibility is an enhanced phonon contribution below $T_{c}$ \cite{Sera}, suppressed by the mixed state. According to the measured phonon contribution to the specific heat and a sound velocity of order 2000~m/s \cite{Maple2003}, a saturation of the phonon mean free path at $\approx$ 20~$\mu$m is required to reproduce the temperature and amplitude of the maximum of the anomaly. This is 10 times smaller than the crystal smallest length, and might come from extended defects in our crystal, like macroscopic voids flux inclusions... But other explanations like a boosted quasiparticle inelastic scattering lifetime (as in the high-$T_{c}$ cuprates \cite{Cohn, Krishana, Yu}) are also possible. In fact, strong enhancement of the microwave conductivity is observed below $T_{c}$, indicating a rapid collapse in quasiparticle scattering \cite{microwave}

In the case of a phonon origin of the anomaly, we expect below 0.3~K  a contribution to $\kappa$ of order  $0.2T^3$~W/K$\cdot$m, which should not change the observed temperature dependence of our thermal conductivity data below 0.2~K:  $\kappa\approx 0.26T^{2}$~W/K$\cdot$m (full line on figure \ref{kappa_all}). In the case of an electronic contribution, it should have disappeared when inelastic scattering is suppressed, which is certainly the case below 0.3~K (see data at 2.5~T). The $T^{2}$-dependence of $\kappa$ down to $T_{c}/40$ indicates low energy quasiparticle excitations. However, it does not fit with the simple theoretical predictions for anisotropic gap with nodes: $\kappa \sim T^{3}$ for line nodes or 2nd order point nodes and $\kappa \sim T^{5}$ for linear point nodes.   It may result from a crossover regime, toward a finite residual value of $\kappa/T$ expected for example in any imperfect sample displaying unconventional superconductivity.  We also note that experimentally the $T^{2}$ dependence of $\kappa$ is  observed in several unconventional superconductors believed to host line nodes, such as CeRIn$_5$(R=Co,Ir) \cite{mov}, Sr$_2$RuO$_4$\cite{suzuki,izawa1}, and CePt$_3$Si \cite{izawa2}.  

%

\begin{figure}[t]
 \includegraphics[width= 15cm]{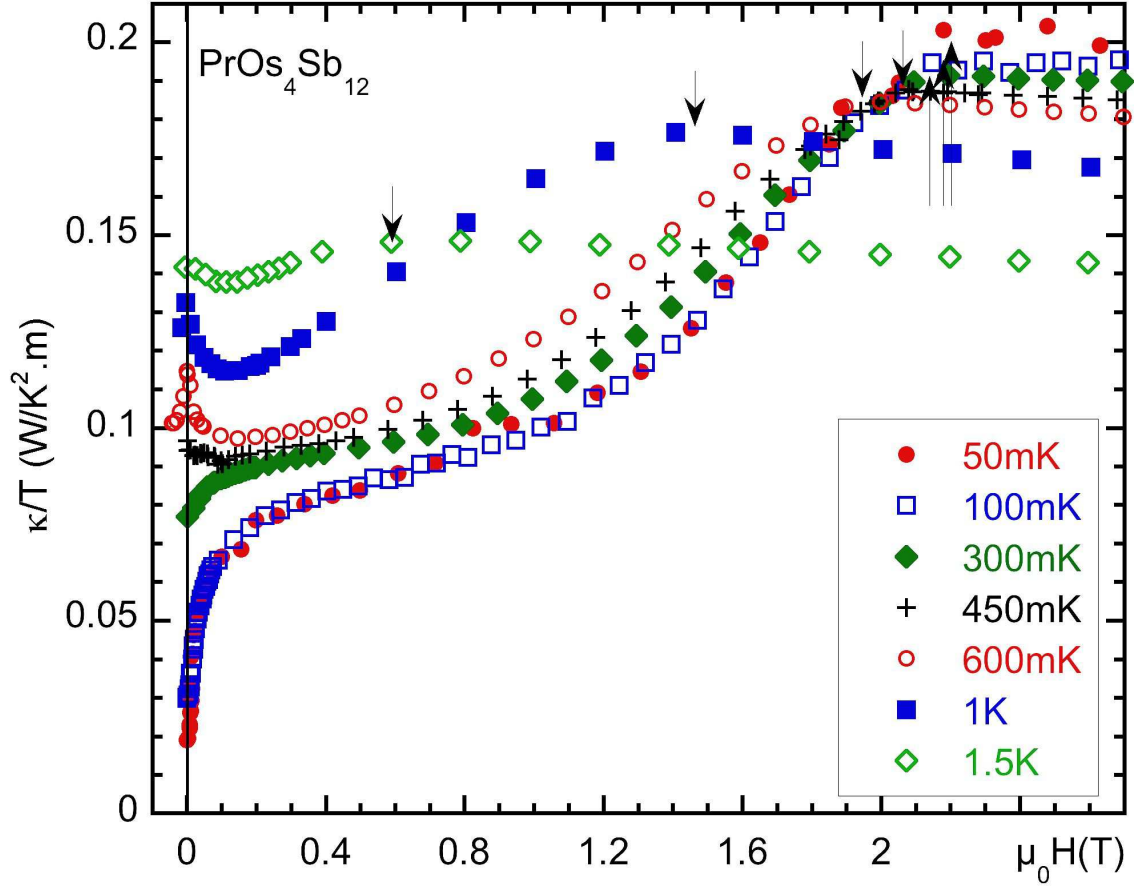}
 \caption{$\kappa(H)$: Field dependence of $\kappa$: around $T_c/2$, it may arise from the strong decrease of the "1K anomaly" (see figure \protect{\ref{kappa_all}}), whereas at low temperatures, the increase signs MBSC (see figure \protect{\ref{compkappa}}). Arrows : $H_{c2}(T)$.}
 \label{kappaH}
 \end{figure}

So let us concentrate on the field dependence $\kappa(H)$ at very low temperatures. As discussed above,  the quasiparticle mean free path below 0.3~K is governed by elastic impurity scattering. We will assume also that the phonon scattering in the same temperature range is governed by static defects and is therefore field independent (at least, it cannot be increased by the field). Under small magnetic fields (20 or 100~mT) at very low temperatures (50 or 100~mK), we observe a pronounced increase in the thermal conductivity with increasing field (figure \ref{kappa_all} and  \ref{kappaH}).  At intermediate fields, we observe a cross-over to a plateau (for $H/H_{c2}\approx$ 0.4) which might be related to the symmetry change observed thermal conductivity experiments under rotating field \cite{Koichi}.

\begin{figure}[t]
 \includegraphics[width=15cm]{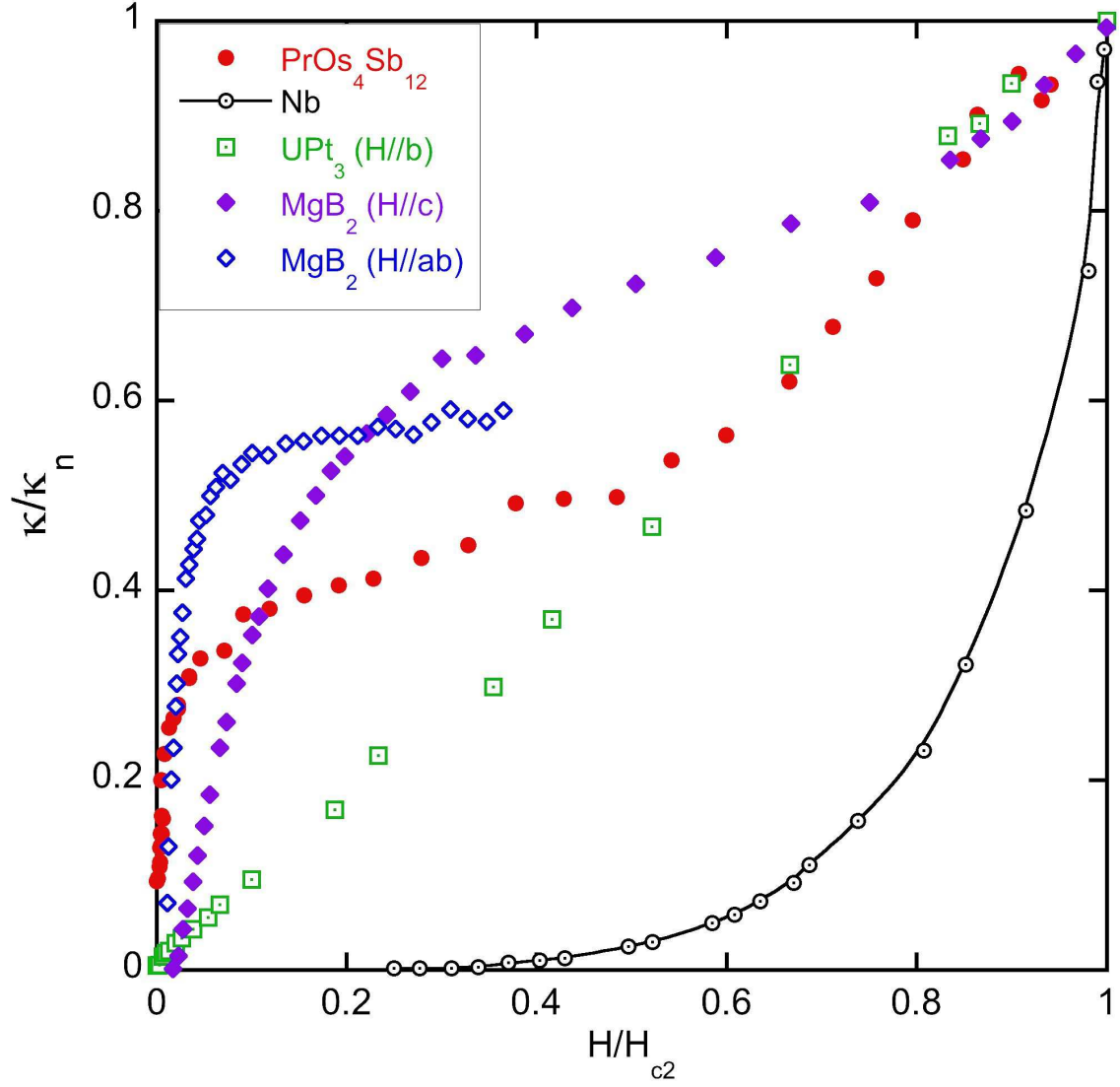}
 \caption{Low temperature behavior of $\kappa(H)$ in a conventional one band superconductors (Nb \cite{Lowell}), UPt$_3$ \cite{Hermann}, MgB$_{2}$ \cite{Sologubenko} and our own data on PrOs$_{4}$Sb$_{12}$: adapted from  \cite{Sologubenko}. The comparison between MgB$_{2}$ and PrOs$_{4}$Sb$_{12}$ is striking, and supports two band superconductivity in this system.}
 \label{compkappa}
 \end{figure}

The $H$-dependence of $\kappa(H)$ in PrOs$_4$Sb$_{12}$ in low fields is in dramatic contrast to that in conventional superconductors.  For conventional superconductors in the clean limit, small magnetic fields hardly affect the very low temperature thermal conductivity. By contrast,  in unconventional superconductors with nodal structure in the gap function, the Doppler shift experienced by the quasiparticles in the mixed state induces a field dependence of $\kappa(H)$.
The initial decrease of $\kappa(H)$ at high temperature (where the condition $\sqrt{H/H_{c2}}<T/T_c$ is satisfied) can be explained by the Doppler shift \cite{Kubert1998}. But the observed $H$-dependence for PrOs$_4$Sb$_{12}$ at low fields is intriguing, since it increases with $H$ steeper than that expected in the Doppler shift \cite{Maki2004}. So, though the Doppler shift can explain the low field behavior of $\kappa(H)$ qualitatively, it is obvious that it cannot explain the whole $H$-dependence.   The extremely strong field dependence of $\kappa(H)$ in PrOs$_{4}$Sb$_{12}$ bears resemblance to that of MgB$_{2}$ \cite{Sologubenko} (see figure \ref{compkappa}). Indeed, half of the normal state thermal conductivity is restored already at $H\approx0.05H_{c2}$ for MgB$_{2}$, and about 40\% of $\kappa$ at $H\approx0.07H_{c2}$ in the case of PrOs$_4$Sb$_{12}$. 
%

As mentioned in the beginning, MgB$_{2}$ is now recognized as the archetype of a two-band superconductor with full gaps, and it is well established experimentally \cite{Bouquet2001, Sologubenko, Giubileo} that the smallest gap on the minor band is highly field sensitive. Theoretically, for $\kappa$, as well as for the specific heat, the field dependence of the smallest gap is controled by a "virtual"  $H_{c2}$ (named $H_{c2}^S$), corresponding to the overlap of the vortex cores of the band with the smallest gap ($\Delta_S$), having a coherence length of order $\frac{\hbar v_F}{\Delta_S}$ \cite{Tewordt2003, Kusunose2002}: 
above  $H_{c2}^S$, the contribution to $\kappa$ of the small band with full gap is close to that in the normal state, {\it only} when it is in the dirty limit (a condition easily satisfied owing to the large coherence length of that band). 
This remains true even if small inter-band coupling prevents a real suppression of $\Delta_S$ at $H_{c2}^S$ \cite{Tewordt2003}. In the case of PrOs$_{4}$Sb$_{12}$, the large ratio of $H_{c2}/H_{c2}^S$ may originate both from the difference in the gap and from the difference in the Fermi velocity between the bands.

However, this "dirty limit" scenario seems incompatible with the unconventional superconductivity revealed by several experiments \cite{Koichi,Chia,Huxley2004}. On the other hand, in case of unconventional superconductivity, MBSC might as well give rise to a rapid increase of $\kappa(H)$ at low temperature even in the clean limit. Indeed, for an unconventional superconductor, one expects an increase of the contribution of the small gap band on a field scale of order $H_{c2}^S$ provided the condition $\sqrt{H/H_{c2}^S}\gg T/T_c$ is satisfied. In PrOs$_4$Sb$_{12}$, this will be the case at $T=0.05~K$ or $0.1~K$ whatever the field above $H_{c1}$ ($\approx2mT$) \cite{Ho2003}.

An additional experimental observation gives support to the MBSC scenario. Indeed, we faced unexpected difficulties in getting reliable measurements in PrOs$_4$Sb$_{12}$ compared to previous works on other systems (see e.g \cite{Hermann}):  it was not until very thin (17~$\mu$m) Kevlar fibers were used for the suspension of the thermometers that reliable values of $\kappa$ (satisfying the Wiedemann-Franz law above  $H_{c2}$) were obtained, and it proved very hard to cool down the thermometers below 30~mK.  Curiously, if a very small field ($\approx$ 10~mT) is applied, the thermometers cool down below 15~mK.  It is nowadays recognized in the community of low temperature thermal measurements, that thermal contacts are a central issue for the reliability of such measurements \cite{Bel}. Before invoking a possible intrinsic mechanism for bad thermal contacts (electron phonon-coupling...), 
we fully characterized these contacts, measuring both their electrical ($R_{c}^{e}$) and the thermal resistance ( $R_{c}^{th}$).
%

\begin{figure}[t]
 \includegraphics[width= 15cm]{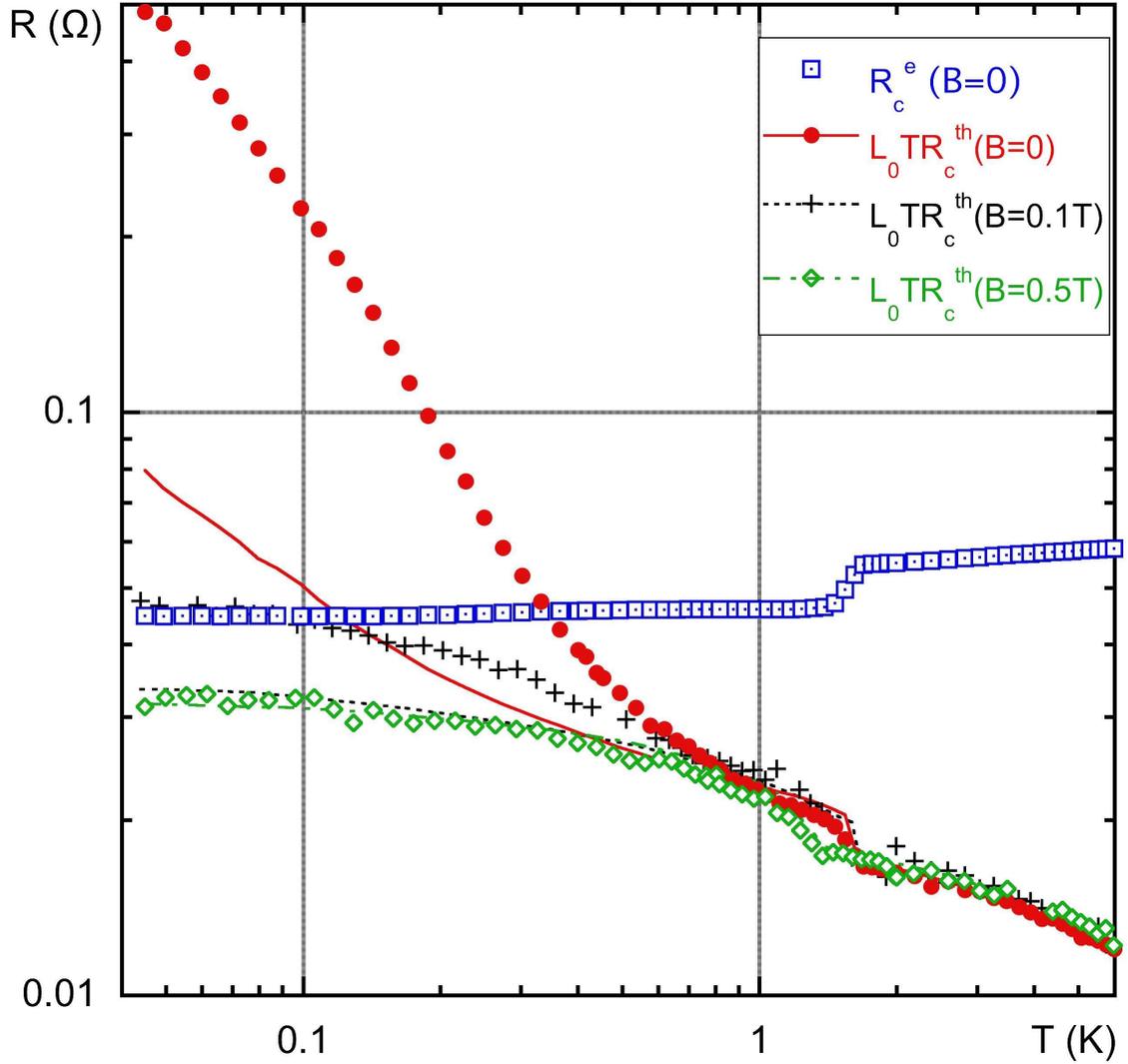}
 \caption{\label{Rcontact}
Squares: $R_{c}^{e}$ of one of the 2 contacts between thermometer and sample (both show the same behavior). By contrast,  $R_{c}^{th}$ (circles, here multiplied by $L_0T$) is strongly diverging at low temperature in zero field, and highly field dependent. Lines are calculated from expression (\protect{\ref{equRcth}}) for each field. On this graph, the thermal leak due to the Kevlar suspension remains always above 10 Ohms.
}
 \end{figure}

The results are shown on Figure~\ref{Rcontact}: despite a constant (Ohmic) $R_{c}^{e}$ below $T_{c}$,  $R_{c}^{th}$ diverges strongly at low temperatures, explaining the difficulties encountered on cooling the thermometers below 30~mK. It is also seen that this divergence is strongly suppressed in a field of 100~mT, much smaller than the upper critical field  $H_{c2}$, in agreement with the observation of the field sensitivity of the base temperature of the thermometers.  $R_{c}^{e}$ has the usual Maxwell contribution (coming from the concentration of current and field lines in the contact area): $R_M^e=\rho/2d$, with $d$ = 17~$\mu$m the gold wires diameter, and $\rho$ the resistivity of PrOs$_{4}$Sb$_{12}$ \cite{Wexler} (we can neglect the resistivity of the gold wires), which controls the temperature dependence of $R_{c}^{e}$ above $T_{c}$ and its jump at $T_{c}$. It also has an additional constant contribution ($R_{cc}\approx $35~m$\Omega$), coming from scattering at the Au-PrOs$_{4}$Sb$_{12}$ interface. We define  $R_{c}^{th}$ as $\Delta T/P$, $P=R_c^e i^2$ the heat power generated by direct joule heating (with current $i$), and $\Delta T$ the thermal gradient across the contact. Following the analysis of $R_{c}^{e}$, $\Delta T$ should have two contributions so that:
\begin{equation}
\label{equRcth}
		R_c^{th}= \frac{R_{cc}}{2R_c^e}\frac{1}{L_0T/R_{cc}+\alpha T^2} + \frac{1}{4d\kappa} (1+\frac{R_{cc}}{R_c^e})\\
\end{equation}

The first term is coming from $\Delta T$ across the interface. We assume a linear increase of the heat power up to $R_{cc}i^2$ within $R_{cc}$, a thermal conduction following the Wiedemann-Franz law for the electronic contribution ($L_0$: Lorentz number), and a $\alpha T^2$ law for the phonon contribution: $\alpha \approx$ 0.18e$^{-6}$~W/K$^3$ is the only free parameter of expression (\ref{equRcth}), and has the same value for all fields. The second term is the Maxwell contribution from the thermal conductivity of the sample ($1/2\kappa d$ \cite{Wexler}), with a uniform heat power ($R_{cc} i^2$) plus a non uniform heat power generated by $R_M^e$ (non zero only above $T_c$) \cite{Wexler}.

Expression (\ref{equRcth}) gives a fair account of  $R_{c}^{th}$ in 500~mT over the whole temperature range, and above 0.8~K in 100~mT and in zero zero field, including the observed jump of  $R_{c}^{th}$ at $T_{c}$ (due to the change in the distribution of heat power when $R_M^e$ is suppressed).
The additional divergence below 0.8~K might well come from the Sharvin surface resistance, which gives additional thermal impedance due to the gap opening even in a metallic contact (Andreev scattering does not contribute to heat transport, \cite{Andreev}). This divergence of   $R_{c}^{th}$ whereas $R_{c}^{e}$ remains ohmic and stable put drastic constraints on the  thermal  insulation of the thermometers from the refrigerator, and can be faced in any other experiments with a constriction at a normal-superconducting interface. On the other hand, the suppression of this divergence in low field in PrOs$_{4}$Sb$_{12}$ is easily explained by the MBSC scenario, because above $H_{c2}^S\ll H_{c2}$, thermal excitations from the normal metal will be transfered in the small gap band without additional barrier. This effect seen on the interface thermal conductivity is even stronger than that observed on the bulk thermal conductivity.

So both $\kappa(H)$ and $R_c^{th}(H)$, which probe the excitation spectrum, give support to multiband superconductivity in PrOs$_{4}$Sb$_{12}$. A possibility for the origin of multiband band superconductivity in this system is the spread in density of states among the various bands of that compound \cite{Marie-Aude}: comparison of de Haas-van Alphen \cite{Sugawara} and specific heat measurements \cite{Bauer I}, reveals that some of them contain quasiparticles with large effective masses ($m^{\ast}\sim50m_{e}$) and the other only light quasiparticles ($m^{\ast}\sim4m_{e}$) \cite{Sugawara}, a situation similar to that of most Ce heavy-fermion compounds. Theoretical work combining band calculations (for the determination of $v_F$) and a realistic fit of $\kappa(H)$ (to extract $H_{c2}^S$) is needed to evaluate precisely the smallest gap. If instead we take the inflection point at low field of $\kappa(H)$ at 50~mK as a "typical value" we get $H_{c2}^S\approx$ 15~mT. With the ratio of the Fermi velocities of both bands extracted from $H_{c2}(T)$ \cite{Marie-Aude}, we find a gap ratio of order 2 : this (rough) estimate contrasts with the very large field effect. It is a direct consequence of the hypothesis that the two bands of the model have very different renormalized Fermi velocities, which may be taken as indicative of weak interband scattering .

So our thermal transport measurements under magnetic field provide clear evidence for MBSC in the HF compound PrOs$_{4}$Sb$_{12}$. Strong electronic correlations are thought to be at the origin of the different coupling between the various electronic bands, as opposed to the difference in the dimensionality of the various sheets of the Fermi surface in MgB$_{2}$.  The low field behavior of $\kappa$ is consistent with unconventional superconductivity. 
Further work on purer samples with improved thermal contact would be very valuable to precise the low temperature behavior of $\kappa(T)$ in this MBSC superconductor in zero field, a study which has not been possible in MgB$_2$ or NbSe$_2$ owing to a dominant phonon contribution.

\begin{acknowledgments}
We are grateful for stimulating discussions with M. Zhithomirsky, H. Courtois,  A. Huxley and H. Suderow.
This work was partly supported by a Grant-in-Aid for Scientific Research Priority Area "Skutterudite" (No.15072206) MEXT, Japan.
\end{acknowledgments}

\end{document}